\documentclass{ws-mpla}

\newcommand{\ket}[1]{| \, #1 \, \rangle}

\begin{document}

\markboth{T. Hyodo, W. Weise, D. Jido, L. Roca, and A. Hosaka}
{$\Lambda(1405)$ in chiral SU(3) dynamics}

\catchline{}{}{}{}{}

\title{$\Lambda(1405)$ IN CHIRAL SU(3) DYNAMICS}

\author{\footnotesize TETSUO HYODO}

\address{Physik-Department, Technische Universit\"at M\"unchen, 
D-85747 Garching, Germany, and\\
Yukawa Institute for Theoretical Physics, Kyoto University, 
Kyoto 606--8502, Japan \\
thyodo@ph.tum.de}

\author{\footnotesize WOLFRAM WEISE}

\address{Physik-Department, Technische Universit\"at M\"unchen, 
D-85747 Garching, Germany}

\author{\footnotesize DAISUKE JIDO}

\address{Yukawa Institute for Theoretical Physics, Kyoto University, 
Kyoto 606--8502, Japan}

\author{\footnotesize LUIS ROCA}

\address{Departamento de F\'{\i}sica. Universidad de Murcia. 
E-30071, Murcia. Spain}

\author{\footnotesize ATSUSHI HOSAKA}

\address{Research Center for Nuclear Physics (RCNP),
Ibaraki, Osaka 567-0047, Japan}

\maketitle

\pub{Received (Day Month Year)}{Revised (Day Month Year)}

\begin{abstract}
We discuss several aspects of the $\Lambda(1405)$ resonance in relation to 
the recent theoretical developments in chiral dynamics. We derive an 
effective single-channel $\bar{K}N$ interaction based on chiral SU(3) 
coupled-channel approach, emphasizing the important role of the $\pi\Sigma$ 
channel and the structure of the $\Lambda(1405)$ in $\bar{K}N$ phenomenology.
In order to clarify the structure of the resonance, we study the behavior 
with the number of colors ($N_c$) of the poles associated with the 
$\Lambda(1405)$, and argue the physical meaning of the renormalization 
procedure.
\end{abstract}

\ccode{PACS Nos.: 13.75.Jz, 14.20.-c, 11.30.Rd, 11.15.Pg}

\section{Introduction}	

The $\Lambda(1405)$ resonance has been drawing intensive attention. On top of
the long-standing problem of its structure (three-quark versus hadronic 
molecule), new issues have been discussed recently: the influence of the 
$\Lambda(1405)$ to the $\bar{K}N$ interaction in relation to the possible 
antikaon bound states in nuclei, and the two-pole structure of the scattering
amplitude for the $\Lambda(1405)$. These issues are stimulated by the recent 
experimental studies as well as theoretical developments of the 
nonperturbative chiral coupled-channel dynamics. Here we address several 
issues of the $\Lambda(1405)$ in connection with the renewed interests on 
this resonance.

\section{Nonperturbative chiral dynamics}	

The low energy hadron physics is systematically studied by chiral 
perturbation theory (ChPT). However, when the hadron interaction is strongly
attractive, we certainly need a nonperturbative technique beyond ChPT. This 
is indeed the case for the $S=-1$ meson-baryon scattering where the leading 
order interaction in ChPT reads
\begin{equation}
    V_{ij}
    \sim -\frac{C_{ij}}{4f^2}(\omega_i+\omega_j) , 
    \label{eq:WT}
\end{equation}
and the diagonal couplings $C_{\bar{K}N}=3$ and $C_{\pi\Sigma}=4$ are 
attractive enough to generate singularities of scattering amplitude. A 
nonperturbative chiral approach has been developed in 
Refs.~\refcite{ChU,Oller:2000fj}, where the coupled-channel scattering 
amplitude $T_{ij}$ is determined by solving the Bethe-Salpeter (BS) equation
\begin{equation}
    T_{ij}
    =V_{ij}+V_{il}G_lT_{lj} ,
    \label{eq:BS}
\end{equation}
with the interaction kernel $V_{ij}$ in Eq.~\eqref{eq:WT}. The amplitude 
constructed in this way reproduces well the $K^-p$ scattering observables, 
with the $\Lambda(1405)$ resonance being generated from the coupled-channel 
meson-baryon dynamics. This framework was applied to the various hadron 
scatterings with the pseudoscalar meson, successfully reproducing 
experimental data of the scatterings and resonance properties. These 
remarkable successes in a variety of channels can be understood that the 
leading order chiral interaction is determined model independently,\cite{WT} 
which is the driving force to generate the resonances.\cite{Exotic}

An interesting observation has been made\cite{Oller:2000fj} that the 
$\Lambda(1405)$ resonance is associated with two poles of the scattering 
amplitude close each other with the same quantum numbers. A simple
explanation of this structure is given in Ref.~\refcite{Hyodo:2007jq}; both 
the $\bar{K}N$ and $\pi\Sigma$ channels are attractive, and each attractive 
interaction provides one singularity. Since the sign and the strength of the 
interaction~\eqref{eq:WT} 
are
determined by the chiral low energy theorem, the
two-pole structure of the $\Lambda(1405)$ is a natural consequence of the 
chiral symmetry in the coupled-channel $\bar{K}N$-$\pi\Sigma$ system.

\section{Effective $\bar{K}N$ interaction based on chiral dynamics}

The study of possible bound state of antikaon in nuclei is a hot 
topic in nuclear physics.\cite{AY} The structure of the $\Lambda(1405)$ is of
great importance to these studies, because the only experimental information 
below $\bar{K}N$ threshold is the spectrum of the $\Lambda(1405)$ in 
$\pi\Sigma$ channel. 

In order to study the few-body nucleus with an antikaon, a realistic 
$\bar{K}N$ potential is needed, which reproduces the scattering amplitude in 
vacuum. In order to derive such a potential based on chiral 
dynamics,\cite{Hyodo:2007jq} we first construct the single-channel $\bar{K}N$
interaction which incorporates the full coupled-channel effects. Next we 
approximate this interaction by a local potential in Schr\"odinger equation, 
keeping the scattering amplitude the same with the prediction of chiral 
dynamics.

An important observation in Ref.~\refcite{Hyodo:2007jq} is that the resonance
structure in the $\bar{K}N$ amplitude appears at around 1420 MeV, not in the 
nominal position of 1405 MeV observed in $\pi\Sigma$ spectrum. The physics 
behind this observation is the strong $\pi\Sigma$ dynamics which eventually
leads to the two-pole structure of the $\Lambda(1405)$. As a consequence of 
the weaker binding energy, the strength of the effective single-channel 
$\bar{K}N$ interaction is roughly one half of the phenomenological 
potential.\cite{AY} The application of this chiral SU(3) potential to the 
three-body $K^-pp$ system\cite{Dote:2008in} provides a smaller binding energy
than the purely phenomenological approach.

It is also found that the local potential approximation works only at around
the threshold and overestimates the amplitude obtained by the BS 
equation~\eqref{eq:BS}, when extrapolated down to $\sqrt{s}<1400$ MeV. This 
indicates the substantial uncertainty in the subthreshold extrapolation of 
the $\bar{K}N$ interaction, which is relevant for the discussion of the 
deeply bound antikaons in nuclei.

\section{Structure of the $\Lambda(1405)$}

One of the recent interests in hadron physics is the structure of the hadron 
resonances. There are several discussions about the structure of the baryon 
resonances: three-quark versus five-quark, or hadronic molecule versus quark 
originated structure. In principle, all these structures eventually stem from
QCD dynamics and mix each other. Nevertheless it helps our physical 
understanding to decompose a resonance state, and inspect the dominant 
component for the resonance. For instance, the $\Lambda(1405)$ can be 
schematically written as
\begin{equation}
    \ket{\Lambda(1405)}
    =N_3\ket{qqq}
    +N_5\ket{qqqq\bar{q}}
    +N_{MB}\ket{B}\ket{M}
    +\dots
    \label{eq:decomp} ,   
\end{equation}
where the third term is understood as the dynamical meson-baryon component
other than the CDD pole,\cite{Castillejo:1956ed} which could be identified 
within the scattering theory of hadrons.\cite{CDDdynamical}
One naively expects that the baryonic resonances 
in chiral dynamics are dominated by this component, but 
this is not always true and substantial CDD pole contribution was found for 
some resonances.\cite{Pelaez:2003dy} Here we attempt to unveil the structure 
in Eq.~\eqref{eq:decomp} by studying the $N_c$ scaling of the $\Lambda(1405)$
poles\cite{Hyodo:2007np} and by utilizing the renormalization 
condition.\cite{Origin}

The study of the $N_c$ behavior is a powerful tool to clarify the quark 
content of hadron resonances,\cite{Pelaez:2003dy} since the $N_c$ scalings 
are known for $\bar{q}q$ mesons and $qqq$ baryons.\cite{Nc} In 
Ref.~\refcite{Hyodo:2007np}, we study the $N_c$ behavior of the 
$\Lambda(1405)$ resonance in chiral dynamics. Because of the nontrivial $N_c$
dependence of the leading order chiral interaction found in 
Ref.~\refcite{Exotic}, the attractive $\bar{K}N$ interaction remains finite
in the large $N_c$ limit. As a consequence, the $\bar{K}N$ bound state exists
in the large $N_c$ limit. The two poles of the $\Lambda(1405)$ behave 
differently from the scaling of ordinary $qqq$ baryons, indicating that the 
$N_3$ component in Eq.~\eqref{eq:decomp} does not dominate the 
$\Lambda(1405)$.

In the framework of nonperturbative chiral dynamics, we propose a method to 
distinguish the dynamical component from the CDD pole contribution by 
studying the physical meaning of the renormalization condition.\cite{Origin} 
We point out that the previous phenomenological chiral models have included 
the CDD pole contribution in the loop function. With a natural 
renormalization condition, we extract the CDD pole contribution hidden in the
loop function into the kernel interaction $V$, as it should be in the 
framework of the $N/D$ method. By examining the phenomenological chiral 
models, we 
find that the amplitude for the $N(1535)$ requires the CDD pole at around
1.7 GeV, while that for the $\Lambda(1405)$ lies at an irelevant energy of 
17 GeV. This implies that the 
$N_{MB}$
component dominates the 
$\Lambda(1405)$, while substantial quark-originate contributions 
$N_3,N_5,\ldots$ are expected in the
$N(1535)$.

\section{Summary}	

The $\Lambda(1405)$ is very unique baryon resonance, and plays an important 
role in various fields of nuclear and hadron physics. The strong $\bar{K}N$ 
interaction is the principle ingredient of the $\Lambda(1405)$, but at the 
same time we should appreciate the strong $\pi\Sigma$ interaction, from the 
viewpoint of chiral symmetry. The two-pole structure of the $\Lambda(1405)$ 
is no longer a theoretical issue of hadron spectroscopy, but is relevant for 
the discussion of $\bar{K}N$ phenomenology. In order to understand the 
structure of the $\Lambda(1405)$, we make use of the $N_c$ scaling and 
renormalization condition. Both the analyses consistently imply that the 
$\Lambda(1405)$ would be dominated by the dynamical content. The precise 
experimental data on $\bar{K}N$ scattering length and $\pi\Sigma$ spectrum 
is highly desired in order to reduce the theoretical uncertainty to explore 
the strongly interacting $\bar{K}N$-$\pi\Sigma$ system, and to clarify the 
structure of the $\Lambda(1405)$ resonance.

\end{document}